\documentclass[12pt, prd, showpacs]{revtex4}
\usepackage{amsmath}
\usepackage{amssymb}

\input{tcilatex}

\begin{document}

\title{Regular black holes with flux tube core}
\author{Oleg B. Zaslavskii}
\affiliation{Astronomical Institute of Kharkov V.N. Karazin National University, 35
Sumskaya St., Kharkov, 61022, Ukraine}
\email{ozaslav@kharkov.ua}

\begin{abstract}
We consider a class of black holes for which the area of the two-dimensional
spatial cross-section has a minimum on the horizon with respect to a
quasiglobal (Krusckal-like) coordinate. If the horizon is regular, one can
generate a tubelike counterpart of such a metric and smoothly glue it to a
black hole region. The resulting composite space-time is globally regular,
so all potential singuilarities under the horizon of the original metrics
are removed. Such a space-time represents a black hole without an apparent
horizon. It is essential that the matter should be non-vacuum in the outer
region but vacuumlike in the inner one. As an example we consider the
noninteracting mixture of vacuum fluid and matter with a linear equation of
state and scalar phantom fields. This approach is extended to distorted
metrics, with the requirement of spherical symmetry relaxed.
\end{abstract}

\keywords{Regular black hole, flux tube, vacuum-like matter}
\pacs{04.70Bw, 04.20.Gz}
\maketitle

%Use showkeys class option if keyword display desired

% It is always \today, today, but any date may be explicitly specified

%\newpage

\section{Introduction}

The long-standing problem of finding regular black holes has different
approaches. Usually, a central region which surrounds such a singularity is
replaced by another one with a regular center due to the choice of special
kind of matter or fine-tuning parameters of the solution \cite{gliner};. for
a review, see \cite{a}. Regular black holes in the theories with
noncommutative geometry were discussed in \cite{non}. Quite different method
was suggested in \cite{mod}, \cite{n} where instead of replacing a singular
centre a regular tubelike inner geometry was used which does not have a
centre at all. In doing so, the outer and inner regions were glued along the
timelike surface which could approach a would-be horizon as closely as one
likes. Actually, such a construction represents not a black hole but a
particular case of so-called quasiblack hole. Such objects, which are also
interesting by themselves, do not contain singularities of curvature but, by
more careful inspection,.reveal another singular features (redshift from the
inner core region becomes infinite, the inner region becomes degenerate with
respect to the time of an external observer, etc. - see \cite{qbh} and
references therein). In the present paper we consider a full-fledged black
hole and show how a completely regular black hole with an inner tubelike
geometry can be constructed.

To this end, gluing is performed along the horizon from the very beginning,
so in general a lightlike (null) shell can arise. To get rid off such a
shell and achieve completely smooth gluing, we select black holes with
special behavior of the near-horizon metric. Namely, we demand that the area
of a spatial cross-section of a constant radius correspond to a local
minimum with respect to a quasiglobal (Kruskal-like) coordinate (see below
for details). It is worth noting that spherically-symmetrical regular
configurations having no center were discussed in \cite{pha1}, \cite{bu} for
the case of phantom scalar fields. In our work, we consider only a quite
definite type of regular black holes but, on the other hand, suggest a quite
general recipe of how to generate them knowing only an original "seed' black
hole that can contain singularities under the horizon.

In the flux tube region, the radial pressure and energy density obey the
equation $p_{r}=-\rho $. This equation proposed for the description of
superhigh density \cite{sakh}, final stage of gravitational collapse \cite%
{gliner} and de Sitter-like core of regular black holes \cite{dym}, appears
in our approach automatically. However, in our approach, the inner core can
have the more general geometry of the type $R_{2}xS_{2}$ where $S_{2}$ is a
sphere of a fixed radius and $R_{2}$ is the two-dimensional de Sitter,
anti-de Sitter or flat spacetime.

\section{Flux tubes and spherically-symmetrical black holes}

Consider an arbitrary spherically-symmetrical black hole. Its metric can be
written in the form%
\begin{equation}
ds^{2}=-fdt^{2}+f^{-1}du^{2}+r^{2}(u)d\omega ^{2}\text{, }d\omega
^{2}=d\theta ^{2}+\sin ^{2}\theta d\phi ^{2}  \label{m}
\end{equation}%
that uses so-called quasiglobal coordinate $u$. It has some advantages in
considering the vicinity of the horizon. In particular, the difference $%
u-u_{h}$ (subscript "h" corresponds to the horizon) is proportional to the
Kruscal-like coordinate \cite{qc}.

The corresponding source is described by the stress-energy tensor%
\begin{equation}
T_{\mu }^{\nu }=diag(-\rho ,p_{r},p_{\perp },p_{\perp }).  \label{set}
\end{equation}

For the metric (\ref{m}) the Einstein equations read

\begin{equation}
G_{0}^{0}=f(\frac{r^{\prime 2}}{r^{2}}+\frac{2r^{\prime \prime }}{r})+\frac{%
f^{\prime }r^{\prime }}{r}-\frac{1}{r^{2}}=-8\pi \rho \text{,}  \label{s00}
\end{equation}%
\begin{equation}
G_{1}^{1}=f\frac{r^{\prime 2}}{r^{2}}+\frac{r^{\prime }f^{\prime }}{r}-\frac{%
1}{r^{2}}=8\pi p_{r}  \label{s11}
\end{equation}%
\begin{equation}
G_{2}^{2}=\frac{fr^{\prime \prime }}{r}+\frac{(r^{2}f^{\prime })^{\prime }}{%
2r^{2}}=8\pi p_{\perp }\text{,}  \label{s22}
\end{equation}%
prime denotes differentiation with respect to $u$.

It is seen from (\ref{s00}) - (\ref{s22}) that there exists a solution with $%
r=const\equiv r_{h}$ provided%
\begin{equation}
\rho =\frac{1}{8\pi r_{h}^{2}}\text{,}  \label{roh}
\end{equation}%
\begin{equation}
p_{r}=-\rho \text{,}  \label{prro}
\end{equation}%
\begin{equation}
f^{\prime \prime }=8\pi p_{\perp }\text{.}  \label{ptr}
\end{equation}

This solution is regular everywhere, provided $p_{\perp }$ is bounded.
Therefore, it can be a candidate inner region of a black hole.

Now we implement the following procedure. We consider a composite spacetime
such that for $u\geq u_{h}$ it has the original form (\ref{m}) while for $%
u\leq u_{h}$ the inner region is replaced by the solution (\ref{roh}) - (\ref%
{ptr}) with $r=r(u_{h})=r_{h}=const$ and constant $\rho $ and $p_{r}$ (\ref%
{roh})- (\ref{prro}). One can see that this is possible if we glue both
regions along the horizon $r=r_{h}$ where $f(r_{h})=0$, and either 1) $%
r_{h}^{\prime }=0$ or 2) $r_{h}^{\prime }\neq 0$, $f_{h}^{\prime }=0$
(extremal horizon). Indeed, it follows from (\ref{s00}) - (\ref{s22}) that
if any of these two conditions is fulfilled, one obtains on the outer side
of the horizon the same relations as on the inner one. In the first case
gluing can be smooth, in the second one a lightlike shell arises on the
horizon. Below we analyze both possibilities separately.

\section{Case 1: completely regular solution}

\subsection{Inner geometry}

Let the condition 
\begin{equation}
r^{\prime }(u_{h}+0)=0  \label{r'}
\end{equation}%
be satisfied. For any regular black hole $\frac{\partial r}{dl}\rightarrow 0$
in the horizon limit where $l$ is the proper distance. However, (\ref{r'})
means a more tight condition which is not satisfied for a generic black
hole. It can be rewritten as 
\begin{equation}
\frac{1}{\sqrt{f}}\frac{dr}{dl}\rightarrow 0  \label{rlf}
\end{equation}%
where on the horizon $f$ itself vanishes.

We want to achieve smooth gluing, so that deltalike singularities should be
excluded. Therefore, the first derivatives of the metric should be
continuous. It follows from (\ref{r'}) and from constancy of $r_{h}$ inside
that $r^{\prime }(u_{h}+0)=r^{\prime }(u_{h}-0)$. Apart from this, we must
also demand that $\frac{df}{du}(u_{h}+0)=\frac{df}{du}(u_{h}-0)$, so $\kappa
_{+}=\kappa _{-}$ where $\kappa $ is the surface gravity. The transversal
pressures $p_{\perp }$.do not necessarily coincide. Moreover, $p_{\perp }$
can be arbitrary bounded function of $u$ inside. Nonetheless, the simplest
choice is to take inside $p_{\perp }=p_{\perp }(u_{h}+0)=const$. This is
physically preferable since it means that we construct a globally regular
solution using only information on the outer side of the horizon and not
inventing equation of state inside \textit{ad hoc}. Then, it follows from (%
\ref{ptr}) that (if $\kappa $ and $p_{\perp }$ do not vanish simultaneously) 
$f=\kappa (u-u_{h})+4\pi p_{\perp }(u-u_{h})^{2}$ in the inner region.
Consider the folllowing possible variants:$:$

a) $p_{\perp }>0$. As in the outer region $f>0$, $p_{\perp }>0$. Thus,
inside there is an inner horizon at $u_{-}=u_{h}-\frac{\kappa }{4\pi
p_{\perp }}$. $\ $We have a nonstatic region for $u_{-}<u<u_{h}$ and a
static one for $u<u_{-}$. \ If $\kappa =0$ (degenerate horizon) $f>0$
everywhere. The four-dimensional inner geometry is AdSxS$_{2}$ where AdS\ is
the two-dimensional space-time, S$_{2}$ is a sphere of a fixed radius. If \
the curvature of both two-dimensional manifolds coincide ($p_{\perp }=\frac{1%
}{8\pi r_{h}^{2}}=-p_{r}$) we obtain the Bertotti-Robinson space-time \cite%
{br}.

b) $p_{\perp }<0$. Then, there is no additional horizon. The geometry is dS$%
_{2}$xS$_{2}$ where dS is the two-dimensional de Sitter space-time. If $%
p_{\perp }=-\frac{1}{8\pi r_{h}^{2}}=p_{r}$ we obtain the Nariai solution 
\cite{nar}.

c) $p_{\perp }=0$. For $\kappa \neq 0$ we have inside $f<0$, so this is the
so-called black universe \cite{bu}. The geometry is R$_{2}$xS$_{2}$ where
now R$_{2}$ is the two-dimensional flat space-time.

The special case arises if $p_{\perp }(u_{h}+0)=0=\kappa $. Then, the
extension of space-time from outside to inside with the constant $p_{\perp
}=0$ is not possible since eq.(\ref{ptr}) is incompatible with the existence
of a horizon in this case. Instead, one should take inside any bounded
function $p_{\perp }(u)$ with $p_{\perp }^{\prime }(u)\neq 0$.

\subsection{Admissible types of outer horizon}

However, this is not the end of the story since now we must formulate more
precisely which regular horizons are compatible with the condition (\ref{r'}%
). We suppose that near the horizon%
\begin{equation}
f=u^{n}F(u)  \label{fn}
\end{equation}%
where $F(u)$ is a sufficiently smooth function, $F(0)<\infty $, $n>0$ is an
integer.

It is also instructive to write the corresponding conditions for the metric
written in the curvature coordinates: 
\begin{equation}
ds^{2}=-fdt^{2}+V^{-1}dr^{2}+r^{2}d\omega ^{2}\text{.}  \label{uv}
\end{equation}%
Let near the horizon%
\begin{equation}
f\sim (r-r_{h})^{q}\text{, }V\sim (r-r_{h})^{p}  \label{fpq}
\end{equation}%
where $q>0$ and $p>0$. Then, it follows from the comparison of (\ref{uv})
and (\ref{m}) that 
\begin{equation}
r-r_{h}\sim u^{k}\text{, }k=\frac{2}{q-p+2}\text{, }
\end{equation}%
\begin{equation}
n=\frac{2q}{q-p+2}  \label{n}
\end{equation}

For (\ref{r'}) to be valid, we must require $k>1$ ($p>q$). We want to have
regular behavior of the metric coefficients and require $k$ to be an
integer, so $k=2,3,...$. Now we can use some previous results concerning
general properties of horizons in spherically-symmetrical space-times \cite%
{hor}. \ Namely, in Table 1 of Ref. \cite{hor} we select those regular cases
which obey the condition (\ref{r'}). Then, we obtain the following set of
possibilities.

\begin{tabular}{|l|l|l|l|l|l|}
\hline
Case & $p$ & $q$ & $n$ & $k$ & Type of horizon \\ \hline
1 & $\frac{3}{2}$ & $\frac{1}{2}$ & $1$ & $2$ & Naked \\ \hline
2 & $\frac{3}{2}<p<2$ & $2-p$ & $1$ & $>2$ & Usual \\ \hline
3 & $\frac{n}{2}+1$ & $\frac{n}{2}$ & $\geq 2$ & $2$ & Naked \\ \hline
4 & $\geq 2$ & $p-2<q<p-1$ & $\geq 2$ & $\geq 2$ & Usual \\ \hline
\end{tabular}

Thus, the regular horizons which obey the condition (\ref{r'}) do exist
including the nonextremal ($n=1$), extremal ($n=2$) and ultraextremal ($n>2$%
) ones. The terms in the last column are explained in the next section.

\subsection{Nature of horizon and possible equations of state}

\subsubsection{Outer region}

In the last column of the table we used classification \cite{tr}, \cite{v},
based on the behavior of curvature tensor in the free-falling frame which
generalizes that suggested in \cite{naked}. Namely, if we denote relevant
combination of curvature component characterizing the magnitude of tidal
forces in the static frame as $Z$ and in the free-falling one as $\tilde{Z}$%
, we have the following classification of cases not containing curvature
singuilarities (see for details \cite{tr}, \cite{hor}): $Z(u_{h})=0=\tilde{Z}%
(u_{h})$ (usual), $Z(u_{h})=0$, $\tilde{Z}(u_{h})\neq 0$ (naked), $\tilde{Z}%
(u_{h})=\infty $ but the Kretschmann scalar remains finite (truly naked).
The latter case is a matter of fact singular (although this is not the
curvature singularity), the metric cannot be extended across the horizon, so
it is now excluded from consideration.

\subsubsection{Inner region}

By itself, the metric (\ref{m})\ with $r=const$ possesses the so-called
acceleration horizon. It arises due to the kinematic effects and can be
removed by the transition to the appropriate frame (see, e.g., the
discussion in \cite{lap}, \cite{hormat}). However, if we consider the
combined metric that contains both the outer region (say, asymptotically
flat region where $f\approx 1-\frac{2m}{u},u\rightarrow \infty $) and the
inner one, the frame is fixed. Then, the signal emitted in the region $%
u<u_{h}$ cannot reach the asymptotic infinity, so there is a true event
horizon. Meanwhile, inside there is no region with ($\nabla r)^{2}<0$ since $%
\left( \nabla r\right) ^{2}=0$ everywhere inside. Therefore, there is no
apparent horizon, so that we have a black hole without an apparent horizon.
It is worth noting that the similar phenomenon was observed for special
types of quasi-black holes \cite{n} (the term "quasi-black hole" was not
used there). Now, it appears\ for a full-fledged black hole.

\subsubsection{Equation of state: outside versus inside}

As far as the properties of matter supporting space-times under
consideration are concerned, it is essential that corresponding black holes
should not be vacuum ones. Indeed, if everywhere on the outer side eq. (\ref%
{prro}) holds, it follows from (\ref{s00}), (\ref{s11}) that $r(u)=u$ and,
thus, eq.(\ref{r'}) cannot be satisfied (it corresponds to $p=q=1$ but there
is no such case in Table 1). On the another hand, in the inner region the
validity of eq. (\ref{prro}) is essential. Thus, it is required that the
equation of state be nonvacuum outside and vacuum-like inside. This is
contrasted with \cite{dym}, \cite{kbmag} where it was assumed that the
source for regular black holes is vacuum-like everywhere.

\section{Examples}

Here we discuss some examples of systems which admit the constructions under
discussion. Although they were also discussed in \cite{hor}, now our goal is
quite different. In \cite{hor}, the conditions for the existence of truly
naked black holes were looked for. Now, we want to analyze, under what
conditions the regular horizons with the additional constraint (\ref{r'})
are possible.

\subsection{Matter with generic linear equation of state}

Consider the case when there is a noninteracting mixture of matter described
by the stress-energy tensor (\ref{set}) and vacuum fluid. By definition,
such a fluid has the stress-energy tensor%
\begin{equation}
T_{\mu (vac)}^{\nu }=diag(-\rho _{(vac)}\text{, }p_{r(vac)}\text{, }p_{\perp
(vac)}\text{, }p_{\perp (vac)})\text{.}  \label{vac}
\end{equation}%
Examples of such vacuum matter can be found in linear and nonlinear
electrodynamics, Yang-Mills theories, for radially-directed cosmic strings 
\cite{sd}, etc. Algebraic structure (\ref{vac}) was used for search of
globally regular solutions \cite{kbmag} and models of dark energy \cite{ki}.

We assume that on the horizon $\rho \rightarrow 0$, $p_{r}\rightarrow 0$.
Then, only the contribution from the vacuum (\ref{roh}), (\ref{prro})
remains there. We also assume that the radial pressure and density are
related (at least, in some vicinity of the horizon) by the linear equation%
\begin{equation}
p_{r}=w\rho 
\end{equation}%
where the parameter $w$ is a constant. Then, it is shown in \cite{cur} that
the horizon is regular provided%
\begin{equation}
w=-\frac{n}{n+2k}.
\end{equation}%
$.$

Now it follows from (\ref{r'}) that we should have $k\geq 2$. Thus, the
parameter $w$ belongs to the range $-1<w\leq -\frac{1}{5}$. For the matter
with such an equation of state regular black holes with flux tube core are
possible. All four cases from Table 1 can be realized.

\subsection{Scalar field with potential}

Let us consider the scalar field $\varphi $ with the potential $V(\varphi )$%
. Then, we can take advantage of the analysis carried out in Sec. V D of 
\cite{hor}. It turns out that the regular horizon is simple ($n=1$), the
power expansion near the horizon has the form 
\begin{equation}
\varphi \approx \varphi _{h}+\frac{\varphi _{1}}{k}(u-u_{h})^{\frac{k}{2}}%
\text{, }r\approx r_{h}+r_{k}(u-u_{h})^{k}\text{,}
\end{equation}%
\begin{equation}
\varphi _{1}\equiv \sqrt{-2\varepsilon k(k-1)r_{k}/r_{h}}\text{,}
\end{equation}%
\begin{equation}
V(\varphi )\approx V_{0}+const(\varphi -\varphi _{0})^{\alpha }\text{, }%
\alpha =\frac{2(k-1)}{k}\text{.}
\end{equation}

Here $\varepsilon =1$ for normal scalar field (with the usual sign of the
kinetic energy) and $\varepsilon =-1$ for the phantom case (negative kinetic
energy). As we want to have a black hole horizon in the region $u>u_{h}$ we
must take $r_{k}>0$, so $\varepsilon =-1$ (the phantom case). To obey the
criterion (\ref{r'}), it is sufficient to take directly $k\geq 2$. Thus,
cases 1 and 2 can be realized (but not cases 3 and 4). It is worth noting
that for odd $k$ the expansion for $\varphi $ contains fractional powers of $%
u-u_{h}$.

It follows from the field equations for the scalar field (see, e.g. \cite%
{pha1} and references therein) that ($fr^{2}\varphi ^{\prime })^{\prime
}=r^{2}\frac{dV}{d\varphi }$. Therefore, in the flux tube region $\varphi
=const=\varphi _{0}$, we obtain $\frac{dV}{\varphi }(\varphi _{0})=0$. The
density $\rho =\frac{V(\varphi _{0})}{8\pi }=const$, $V(\varphi _{0})=\frac{1%
}{r_{h}^{2}}$ in accordance with (\ref{roh}).

\section{Generalization to distorted black holes}

In this section we generalize the above results to the distorted case
relaxing the requirement of spherical symmetry. In the Gauss normal
coordinates any static metric can be written (at least, in some region) as

\begin{equation}
ds^{2}=-dt^{2}N^{2}+dl^{2}+\gamma _{ab}dx^{a}dx^{b}{,}  \label{md}
\end{equation}%
where $x^{1}=l$, $a=2,3$. Then, using 2+1+1 decomposition, one can write the
Einstein tensor in the form \cite{vis}%
\begin{equation}
G_{0}^{0}=-\frac{1}{2}R_{\parallel }-\frac{\partial K}{\partial l}+\frac{1}{2%
}K_{ab}K^{ab}+\frac{1}{2}K^{2}  \label{g00}
\end{equation}%
\begin{equation}
G_{l}^{l}=-\frac{1}{2}R_{\parallel }-\frac{1}{2}K_{ab}K^{ab}+\frac{\Delta
_{2}N}{N}-K\frac{\partial N}{\partial l}
\end{equation}%
\begin{equation}
G_{l;a}=K_{;a}-K_{ab}^{;b}-K_{a}^{b}\frac{_{N_{;b}}}{N}-\frac{1}{N}\frac{%
\partial N_{;a}}{\partial l}
\end{equation}%
\begin{equation}
G_{ab}=-\frac{N_{;a;b}}{N}+\frac{K_{ab}}{N}\frac{\partial N}{\partial l}+%
\frac{\partial K_{ab}}{\partial l}+2K_{ac}K_{b}^{c}+\gamma _{ab}X
\end{equation}%
\begin{equation}
X=\frac{\Delta _{2}N}{N}+\frac{1}{N}\frac{\partial ^{2}N}{\partial l^{2}}-K%
\frac{\partial N}{\partial l}-\frac{\partial K}{\partial l}+\frac{1}{2}%
K_{ab}K^{ab}\text{.}  \label{x}
\end{equation}%
Here $K_{ab}=-\frac{1}{2}\frac{\partial \gamma _{ab}}{\partial l}$ is the
extrinsic curvature tensor of the surface $l=const$ embedded into the
three-dimensional space$.$

The flux tube solution which generalizes that with $r=const$ for the metric (%
\ref{m}) is defined according to 
\begin{equation}
K_{ab}=0,  \label{k0}
\end{equation}%
so that $\gamma _{ab}=\gamma _{ab}(x^{2},x^{3})$ \cite{kf}. Then, it follows
from (\ref{md}) - (\ref{x}) that there exist a solution with $N=N(l)$
provided the stress-energy tensor of the source in coordinates (\ref{md})
has the same form (\ref{set}) and 
\begin{equation}
\rho +p_{\parallel }=0\text{,}  \label{pr}
\end{equation}%
\begin{equation}
R_{\parallel }=16\pi \rho \text{,}  \label{r2}
\end{equation}%
\begin{equation}
N^{-1}\frac{d^{2}N}{dl^{2}}=\frac{1}{2}\frac{d^{2}N^{2}}{du^{2}}=8\pi
p_{\perp }\text{,}  \label{pt}
\end{equation}%
where $R_{\parallel }$ is the Riemann curvature of the two-dimensional
cross-section $t=const$, $l=const$, $du=dlN$. If $p_{\perp }$ is finite, the
solution is regular.

Let us suppose that we have a black hole solution (which, in general,
contains singularities under the horizon) and want to construct a composite
solution by replacing the region under the horizon by the flux tube. To
achieve smooth gluing, we must generalize the condition (\ref{r'}). As now
we have (\ref{k0}) from inside, it would seem that it is sufficient to
require the same equality from outside. This is not so. For any horizon the
condition of the finiteness of the Kretschmann scalar entails (\ref{k0}) but
for smooth gluing with a flux tube we need something more. It is easy to
understand that the corresponding condition can be written as%
\begin{equation}
\frac{K_{ab}}{N}\rightarrow 0  \label{kn}
\end{equation}%
on the horizon that is generalization of eq. (\ref{rlf}). Indeed, for the
spherically-symmetrical case it reduces to (\ref{r'}), (\ref{rlf}) directly.
For a general case, it is convenient to rewrite the metric (\ref{md}) with
the help of the quasiglobal coordinate $u$ now defined according to%
\begin{equation}
ds^{2}=-N^{2}dt^{2}+\frac{du^{2}}{N^{2}}+\tilde{\gamma}_{ab}dx^{a}dx^{b}%
\text{.}  \label{ud}
\end{equation}%
Then, the condition (\ref{kn}) can be rewritten as $\frac{\partial \gamma
_{ab}}{\partial u}\rightarrow 0$ and analogy with (\ref{r'}), (\ref{rlf})
becomes transparent.

One reservation is in order. In general, when the lapse function depends on
all three coordinates $u,x^{2},x^{3}$ the metrics $\tilde{\gamma}_{ab}\neq
\gamma _{ab}$. However, near the horizon the dependence of $N$ on $x^{a}$
can be neglected. Indeed, let us again assume that, in accordance with (\ref%
{fn}) $N\sim (u-u_{h})^{\frac{n}{2}}$ near the horizon. Then, the regularity
of the metric entails that for the nonextremal horizons ($n=1$) the
finiteness of the Kretschmann scalar leads to the validity of expansion \cite%
{vis}%
\begin{equation}
N=\kappa l+\kappa _{3}(x^{2},x^{3})l^{3}+o(l^{3})  \label{1}
\end{equation}%
where the surface gravity $\kappa =const$. For extremal horizons ($n=2$)%
\begin{equation}
N=B_{1}\exp (-bl)+B_{2}\exp (-2bl)+...\text{, }l\rightarrow \infty \text{,}
\label{2}
\end{equation}%
and \ for ultraextremal horizons ($n\geq 3$) 
\begin{equation}
N=C_{n}l^{-\frac{n}{n-2}}+C_{n+2}l^{-\frac{n+2}{n-2}}+...  \label{3}
\end{equation}

It was shown that for regular horizons the coefficients $B_{1}$, $B_{2}$, $%
C_{n}$, $C_{n+2}$ are constants \cite{v}, \cite{tr}. Thus, in the main
approximation the dependence of $N$ on $x^{a}$ does indeed drop out. Then,
the transition from (\ref{md}) to (\ref{ud}) near the horizon can be made by
a simple substitution $du=Ndl$ and $\tilde{\gamma}_{ab}\approx \gamma _{ab}$.

Now let us take into account the condition (\ref{kn}) in Einstein equations.
Then, for all three types of the horizons (\ref{1}) - (\ref{3}) we obtain
the same equations (\ref{pr}) - (\ref{pt}) as for the flux tubes. For the
metric $\gamma _{ab}$ the corresponding expansion near the horizon should
have the form%
\begin{equation}
\gamma _{ab}\approx \gamma _{ab}(x^{2},x^{3})+\gamma _{ab}^{(k)}(u-u_{h})^{k}%
\text{.}
\end{equation}%
Here $k>1$. As we are interested in the regular case, the quantity $k$
should be integer, so that $k\geq 2$.

It was observed in \cite{gu} that for nonspherical topology of \ a horizon
new possible types of equilibrium between matter and a horizon can arise
which are absent for the spherical one. This is also true for composite
space-times, if the analogues of the Bertotti-Robinson or Nariai space-times
have the corresponding topology. The analysis can also be extended to a
higher-dimensional case \cite{vit}. However, more detailed treatment of
these issues is beyond the scope of the present paper.

\section{Case 2: gluing across null shell - mass without mass}

In the above consideration, we considered globally regular solutions. This
implied that (i) there are no singularities under the horizon, (ii) gluing
between the outer black hole region and the inner flux tube is smooth, so
there is no shell between them on the horizon. Meanwhile, the deltalike
shells is quite legitimate object in general relativity, both in its
timelike or spacelike version \cite{isr} and lightlike one \cite{bisr}, \cite%
{bh}, \cite{pois}. Therefore, we may relax condition (ii) leaving (i)
intact. Although such a solution is not completely regular, it has some
limited interest to be discussed briefly below. One of motivations comes
from the problem of finding self-consistent analogue of the Abraham-Lorentz
electron in general relativity (see, e.g. \cite{vf}). Usually, it is
required that in the pure field model electromagnetic repulsion be balanced
by gravitation, other sources (bare stresses) being absent, the mass having
pure electromagnetic origin. However, the distinguished role of a horizon
opens here an additional possibility. It was observed for quasiblack holes
that in the extremal case the contribution of the surface stresses (which by
themselves do not vanish in general) tends to zero when the shell between
the outer and inner regions approaches the quasihorizon \cite{mod}, \cite{n}%
, \cite{mass}. As a result, the requirement can be weakened. Namely, even
with nonzero bare stresses the ADM mass measured at infinity \cite{Arn} can
be of pure electromagnetic origination.

Instead of a quasiblack hole one can take a full-fledged black hole (say,
the Reisnerr-Nordstr\"{o}m one) and glue it to the inner Bertotti-Robinson
solution for which $r=const$. As for the Bertotti-Robinson space-time $%
r_{h}=e$ ($e$ is an electric charge), only the extremal Reissner-Nordstr\"{o}%
m space-time is suitable for gluing. Using the methods described in detail
in \cite{bisr} and the monographs \cite{bh} [especially see Sec. V and eq.
(5.37) there], \cite{pois} with slight modification (now $r=const$ inside,
so it cannot be taken as an independent variable) one can obtain the
effective density $\mu \,$\ and pressure $P$ of the shell. Taking for
simplicity the extremal horizon also inside, one immediately obtains \cite%
{bh} that $P=0$. However, because of the fact that first derivatives of the
angular part of the metric on both sides of the horizon do not coincide,
gluing is not smooth and the energy density $\mu =-\frac{1}{4\pi r_{h}}$does
not vanish (cf. with \cite{bi} where another result was obtained).

The fact that $\mu <0$ somewhat restricts physical significance of the
composite space-time but, nonetheless, it can be served as an example of
pure electromagnetic mass in the presence of bare stresses in addition to
previous observation about quasiblack holes \cite{mod}, \cite{mass}. Since
there are no singular sources inside, electromagnetic field extends to
infinity inside. Although the null shell arises on the horizon inevitably,
such a shell leaves the gravitational mass across the shell continuous \cite%
{bisr}, \cite{bh}, even irrespective of whether the horizon is extremal or
nonextremal. Thus, the presence of the shell in this case can also be
compatible with the idea of a pure electromagnetic "mass without mass" \cite%
{wh} measured by a remote observer in the outer region.

\section{Summary and conclusion}

We developed a simple and rather general approach that enables, knowing only
the properties of the black hole metric in the outer region, to construct a
completely regular composite space-time with a tubelike geometry under the
horizon. Actually, in this way one can generate the inner regular part of
the metric from the outer vicinity of the horizon only. The area of
applicability of the approach is restricted by requirement (\ref{r'}) or (%
\ref{kn}) that includes, however, sufficiently wide class of physically
relevant systems. In doing so, we do not require special types of the
equation of state inside, it turns out to be the vacuumlike one
automatically. We obtained also the metrics which look as black holes for a
distant observer but have the character of black universes inside \cite{bu}.
One can also glue a black hole on both sides of a flux tube and obtain a
wormhole with a tunnel of arbitrary length between them. This wormhole is,
however, not traversable since horizons are present.

Tube-like geometries discussed in the present work are encountered not only
in general relativity but in more complicated theories of gravity as well
(in particular, in quadratic gravity \cite{jd}), where they can be also used
for constructing regular black holes. They can be relevant also for the
higher-dimensional case \cite{vit} including the Kaluza-Klein theories where
such geometries can arise as compactified phases inside non-compactified
ones \cite{gl}, \cite{flux}. The flux tubes appearing in the space-times
considered in our work, can be also thought of as a special kind of
non-empty "voids" inside matter filled with vacuumlike fluid (cf. \cite{kr}%
). For extrenal observers, they appear as bolls of finite radius and mass
binding an infinite proper mass similarly to what happens to T-models \cite%
{rub}. But, in contrast to T-models, now the space-times under discussion
can be everywhere static if the order of the horizon $n\geq 2$.

In the present work, we restricted ourselves by static space-times. Of
especial interest is to extend the present results to the rotating and
higher-dimensional space-times that needs separate treatment.


\begin{thebibliography}{99}
\bibitem{gliner} E. B. Gliner, Sov. Phys. JETP \textbf{22}, 378 (1966).

\bibitem{bard} J. Bardeen, in GR 5 Proceedings, Tbilisi, 1968.

\bibitem{fmm} V. P. Frolov, M. A.\ Markov and V. F. Mukhanov, Phys. Rev. D%
\textbf{\ 41} (1990) 383.

\bibitem{dym} I. Dymnikova, Gen. Rel. Grav. \textbf{24}, 235 (1992); Phys.\
Lett. B \textbf{472}, 33 (2000); Int.J.Mod.Phys. D \textbf{12,} 1015 (2003).

\bibitem{em} E. Elizalde and S. R. Hildebrandt, Phys. Rev.\textbf{\ D} 65
124024 (2002).

\bibitem{bur} A. Burinskii, E. Elizalde, S. R. Hildebrandt and G. Magli,
Phys. Rev. D \textbf{65}, 064039 (2002).

\bibitem{a} S. Ansoldi, arXiv:0802.0330.

\bibitem{non} P. Nicolini, Int. J. Mod. Phys. A 24, 1229 (2009).

\bibitem{mod} O. B. Zaslavskii, Phys. Rev. D \textbf{70} (2004) 104017.

\bibitem{n} O.B.Zaslavskii, Phys.Lett. \textbf{B634} (2006) 111.

\bibitem{qbh} J. P. S. Lemos and O B. Zaslavskii, Phys. Rev. D \textbf{76},
084030 (2007).

\bibitem{pha1} K.A. Bronnikov and J.C. Fabris, Phys. Rev. Lett. \textbf{96}
(2006) 251101.

\bibitem{bu} K.A. Bronnikov, N. Dehnen and V. N. Melnikov, Gen.Rel.Grav.%
\textbf{39}, 973 (2007).

\bibitem{sakh} A. D. Sakharov, Sov. Phys. JETP \textbf{22}, 241 (1966)..

\bibitem{qc} K. A. Bronnikov, Phys. Rev. D \textbf{64}, 064013 (2001); K. A.
Bronnikov, G. Cl\'{e}ment, C. P. Constantinidis, and J. C. Fabris,
Gravitation Cosmol. \textbf{4}, 128 (1998); Phys. Lett. A 243, 121 (1998).

\bibitem{nar} H. Nariai, Sci. Rep. Tohoku Univ., Ser. 1 \textbf{34}, 160
(1950); 35, \textbf{62 }(1951).

\bibitem{hor} K A Bronnikov, E. Elizalde, S. D. Odintsov and O B Zaslavskii,
Phys. Rev. D \textbf{78}, 064049 (2008).

\bibitem{sd} P. S. Letelier, Phys. Rev. D \textbf{20}, 1294 (1979).

\bibitem{kbmag} K A Bronnikov, Phys. Rev. D \textbf{63}, 044005 (2001).

\bibitem{ki} K. A. Bronnikov and I. G. Dymnikova, Classical Quantum Gravity
24, 5803 (2007).

\bibitem{br} B. Bertotti, Phys. Rev. \textbf{116} (1959) 1331; I. Robinson,
Bull. Acad. Pol. Sci. \textbf{7} (1959) 351.

\bibitem{tr} O.B. Zaslavskii, Phys. Rev. D \textbf{76}, 024015 (2007).

\bibitem{v} V. Pravda and O. B. Zaslavskii, Class. Quant. Gravity, \textbf{22%
}, 5053 (2005).

\bibitem{naked} G. T. Horowitz and S. F. Ross, Phys. Rev. D \textbf{56},
2180 (1997); ibid. \textbf{57}, 1098 (1998).

\bibitem{cur} K A Bronnikov and O B Zaslavskii, Phys. Rev. D \textbf{78},
021501(R) (2008).

\bibitem{kf} O.B. Zaslavskii, Class. Quantum Gravity, \textbf{23}, 4083
(2006).

\bibitem{lap} A. S. Lapedes, Phys. Rev. D \textbf{17} (1978) 2556.

\bibitem{hormat} O.B. Zaslavskii, Class. Quant. Gravity, 1998, \textbf{15},
3251 (1998).

\bibitem{vis} A. J. M. Medved, D. Martin and M. Visser Class, Quantum
Gravity, \textbf{21} 3111 (2004).

\bibitem{gu} K. A. Bronnikov and O. B. Zaslavskii, arXiv:0904.4904 (To
appear in Class. Quant. Gravity.)

\bibitem{isr} W. Israel, Nuovo Cimento B \textbf{44}, 1 (1966).

\bibitem{bisr} C. Barrab\`{e}s and W.\ Israel, Phys. Rev. D \textbf{43},
1129 (1991).

\bibitem{pois} E. Poisson, \textit{A relativist's toolkit}. Cambridhe
University Press, 2004.

\bibitem{bh} C. Barrab\`{e}s and P. A. Hogan, \textit{Singular Null
Hypersurfaces in General relativity}. World Scientific Publishing Co. Pte.
Ltd. 2003.

\bibitem{vit} V. Cardoso, O. J.C. Dias, and J. P.S. Lemos, Phys.Rev. D 
\textbf{70} 024002 (2004).

\bibitem{mass} J. P. S. Lemos and O. B. Zaslavskii, Phys. Rev. D \textbf{78}
(2008) 124013 .

\bibitem{Arn} R. Arnowitt, S. Deser, and C. W. Misner, Phys. Rev. \textbf{120%
}, 321 (1960).

\bibitem{bi} S. H. Mazharimousavi and M. Hallisoy, Phys. Lett. \textbf{B678}%
, 407 (2009).

\bibitem{wh} J. A. Wheeler, \textit{Geometrodynamics }(Academic Press, New
York 1962).

\bibitem{vf} A. V. Vilenkin and P. I. Fomin, Nuovo Cimento Soc. A \textbf{45}%
, 59 (1978).

\bibitem{jd} J. Matyjasek and D. Tryniecki, Phys. Rev. D \textbf{69}, 124016
(2004).

\bibitem{gl} Guendelman, Gen. Rel. Grav., \textbf{23} (1991) 1415.

\bibitem{flux} V. Dzhunushaliev, U. Kasper and D. Singleton, Phys.Lett. B 
\textbf{479 }(2000) 249; V. Dzhunushaliev, Int.J.Mod.Phys. D\textbf{14 }%
(2005) 1293.

\bibitem{kr} A. Krasi\'{n}ski and C. Hellaby, Phys. Rev. D \textbf{69},
023502 (2004).

\bibitem{rub} V. A. Ruban, Soviet Physics JETP \textbf{29}, 1027 (1969)
(Reprinted in Ge. Rel. Grav. \textbf{33}, 375 (2001).
\end{thebibliography}
\end{document}